# A full superconducting dome of strong Ising protection in gated monolayer $WS_2$

J. M. Lu[1], O. Zheliuk[1], Q. H. Chen[1,2], I. Leermakers[3], N. E. Hussey[3], U. Zeitler[3], and J. T. Ye[1*]

[1] *Device Physics of Complex Materials, Zernike Institute for Advanced Materials,*

*Nijenborgh 4, 9747 AG, Groningen, The Netherlands*

[2] *Department of Physics, Hong Kong University of Science and Technology,*

*Clear Water Bay, Hong Kong, China*

[3] *High Field Magnet Laboratory (HFML-EMFL), Radboud University,*

*Toernooiveld 7, 6525 ED Nijmegen, The Netherlands*

**Many recent studies show that superconductivity not only exists in atomically thin monolayers but can exhibit enhanced properties such as higher transition temperature and stronger critical field. Nevertheless, besides being air unstable, weak tunability in these intrinsically metallic monolayers has posed a severe limitation in exploring monolayer superconductivity, hence hindering possible applications in electronic devices. Using field effect gating, we prepared monolayer superconductivity in easily accessible CVD-grown $WS_2$, a typical ambient stable semiconducting transition metal dichalcogenide (TMD). Here, we present a complete set of competitive electronic phases over an unprecedented range from band insulator, superconductor, to an unexpected re-entrant insulator. Throughout the superconducting dome, the Cooper pair spin is pinned by a strong internal spin-orbit interaction, making this material arguably the most resilient superconductor in a magnetic field. The re-entrant insulating state at positive high gating voltages is plausibly attributed to localization induced by the characteristically weak screening of the monolayer, providing the key insight into many dome-like superconducting phases observed in field induced quasi-2D superconductors.**

*email address: j.ye@rug.nl

Reducing the dimensionality from three (3D) to two (2D) often conserves the fundamental electronic properties while simplifies theoretical approaches, making 2D a fruitful dimensionality allowing exact solutions of many important physical models[1,2]. Physically, although quantum confinement increases gradually with reduced thickness, significant modification of electronic states is often observed when approaching the monolayer limit: *e.g.* producing a massless Dirac band in graphene[3] and changing the band gap from indirect to direct in $MoS_2$[4]. On the other hand, the dimensionality of a system is not an invariable but rather related to specific microscopic regimes[5] since it is determined by comparing the geometrical thickness $d$ with fundamental electronic length scales such as the phase coherence length, the Fermi wavelength $\lambda_F$ and the superconducting coherence length $\xi$. For instance, 2D superconductivity is well established in amorphous films of superconducting metals[6] far thicker than a single atom because $\xi$ can easily exceed $d$. However, due to large carrier density in metals, $\lambda_F$ of diffusive electrons appearing after a quantum phase transition is typically only a few angstroms, which is usually smaller than $d$. This means that the transition is associated with a dimensional crossover from a 2D superconductor to a quasi-2D diffusive system. Similarly, this dimensional crossover has been widely observed in interface superconductors and cuprates approaching the optimal gating and doping, respectively[7,8].

These imposed requirements have motivated the search of truly monolayer superconductors. Recently, epitaxial growth on optimized substrates for single-atomic thick metal films[9–11] (Pb, In, Ga *etc.*), monolayer FeSe[12], monolayer cuprates[13,14], Heavy Fermion[15] systems and organic superconductors[16] can satisfy the requirements whereas they show strong interaction with the substrates, making the truly 2D systems still in debate. In particular, the electronic and vibrational couplings in the third dimension are responsible for a reduced critical temperature $T_c$ in metallic monolayers and a significant enhanced $T_c$ in FeSe, respectively. Naturally, van-der-Waals layered materials[17,18] are 2D systems where electrons are mainly confined in a covalently bonded crystalline plane. Therefore, by breaking the van-der-Waals stacking, monolayer superconducting TMD (such as $NbSe_2$[19,20]) and high-temperature superconductor cuprate[14] could exhibit truly 2D characteristics.

However, due to the large intrinsic carrier density, switching electronic phases in these superconductors appears difficult[21,22].

Here we demonstrate that semiconducting TMD monolayers $WS_2$, where both carrier tunability and true 2D characteristics are accessible, provide the unique all-round option for field effect control of various quantum phases. By field effect gating, $WS_2$ flakes evolve from a direct band insulator to metallic states, which develop into Ising superconductivity[20,23,24] at low temperatures. The significant spin-orbit coupling in the conduction band leads to the robust Ising superconductivity against external in-plane magnetic fields. Beyond the peak of the superconducting dome, with increasing gate bias normal state becomes more resistive and insulating behavior emerges, which grows to be more pronounced and eventually quenches the superconducting dome.

Monolayer $WS_2$ is grown by chemical vapor deposition. The high quality of the as-grown crystalline sheets is confirmed by strong photoluminescence (PL) and mobility of $\mu \sim 300$ cm$^2$/Vs at 10 K and the optimal gating, which are comparable or higher than cleaved counterparts[25] (Fig. S1). An electronically homogeneous part showing a uniform PL is isolated by etching a standard Hall-bar from a pristine triangular monolayer (see Method) and we fabricated a device with a dual gate configuration composed of an ionic liquid as a top gate ($V_{TG}$) and a dielectric back gate (285 nm $SiO_2$, $V_{BG}$) as shown in Fig. 1a. This dual-gating method allows the coarse and fine electrostatic control of quantum phases: applying firstly a maximal ionic gate at $T \sim 220$ K we can introduce the strongest electrostatic field effect, which is subsequently fixed by freezing the ionic liquid below its glass transition temperature $T_g \sim 190$ K. By grounding the ionic gate and sequentially warming the device slightly above $T_g$, we are then able to create dense coarse doping states (colour coded in Fig. 1b), which can then be seamlessly linked at low temperatures by $V_{BG}$. The quasi-continuous transfer curve thus obtained (black curve in Fig. 1b) determines electronic phases as a function of effective gate voltage $V_{eff}$ (Methods). Note that on the two sides of the optimal doping, the dependence of Hall carrier density on the $V_{eff}$ are opposite: positive on the left while negative on the right. Around optimal doping, there is a crossover linking two monotonic correspondences (as elaborated in Fig. S2 and S3). Therefore, we use $V_{eff}$ instead of Hall carrier density for labeling states for the whole spectrum of their variation. Although a simple capacitance model cannot be applied straightforwardly in our dual gated device[26], the $V_{eff}$ generally corresponds to the charge carriers injected into the sample, which is equivalent to the nominal stoichiometry of intercalated/doped bulk compound. The $V_{eff}$ dependence of the square resistance $R_s$ (Fig. 1b) at 2 K (diamond), 10 K (triangle), 70 K (square), and 150 K (circle) gives an overview of the whole spectrum of electronic states. Amid two insulating phases, metallic transport appears at medium $V_{eff}$ with a superconducting ground state. At $V_{eff}$ = 1130 V, optimal transport shows the lowest $R_s$ and the highest superconducting transition temperature $T_c$.

The sheet resistance $R_s$ as a function of temperature at the left and the right sides of optimal doping is shown in Fig. 1c and d, respectively. The curves show that superconducting ground states emerge from an insulating state ($dR_s/dT < 0$, Fig. 1c) when approaching optimal doping from low gating and quench into another re-entrant insulating state at high gating (Fig 1d). This exemplifies how a full spectrum of electronic phases can be prepared using a single tuning parameter $V_{eff}$, where properties of quantum phases such as quantitative dependence of superconductivity on the $V_{eff}$ (insets in Fig. 1c and d) can be analyzed in detail.

The critical temperature $T_c$ is determined by widely used 50% of normal state resistance. A full phase diagram is shown in Fig. 2, with a superconducting dome ranging from $V_{eff}$ = 0.8 to 1.6 kV. The $V_{eff}$ dependence of Hall carrier density $n_{Hall}$ (lower panel, Fig. 2) show a clear correlation to the superconducting dome indicating $T_c$ is mainly driven by $n_{Hall}$. Nevertheless, the discrepancy in $V_{eff}$ between maxima of $T_c$ = 4.15 K and $n_{Hall}$ (10 K) = 3.3×10$^{13}$ cm$^{-2}$ may indicate $T_c$ is affected by electron-impurity scattering rate as well which can be inferred from the opposite $V_{eff}$ dependence of carrier mobility $\mu$ ($T$ = 10 K, right axis in lower panel of Fig.2) extracted from $R_s$ (Fig.1b) below and above the $T_c$ maxima. Notably in MoS$_2$[27], the similarly gated superconducting dome shows $T_c \propto (n_{2D} - n_0)^{zv}$, where $zv \sim 0.6$ are the product of exponents for correlation length and correlation time in scaling theory[5,28,29], in analogy to that for LaSrO$_3$/SrTiO$_3$ interfaces[7]. In the present monolayer superconductor, limited temperature range cannot resolve the scaling exponents and exact locations of two quantum critical points (QCPs, denoted roughly by two dashed circles); nevertheless, the voltage dependence of $T_c$ does not contradict the relationship in the above two systems.

In many quasi-2D superconductors[7,30,31] quasi-metallic (qM) regions are speculated[5], in which $R_s$ shows temperature dependence weaker than exponential although overall $dR_s/dT < 0$. Here, we define $T_{min}$ as temperature when $R_s$ reaches a minimum (Fig. 1c, d) plausibly resulting from competition between electron-phonon (*e*–*ph*) scattering and Anderson localization. In Fig. 2, the trace of $T_{min}$ (empty square, upper panel) marks the boundary of the qM regime. It is worth noting that the qM region here is identified from a finite-size sample at finite temperatures which might crossover to an insulator at lower temperatures and/or in larger samples[7]. We leave this question open to the future experiments at ultralow temperatures.

The unique feature of 2D superconductivity in monolayer WS$_2$ is the Ising pairing[20,23,24] originated from valley coupled spin texture of monolayer TMDs. Here, the spins in Cooper pairs are pinned by strong out-of-plane effective magnetic field, originated from an intrinsic Zeeman type spin-orbit interaction (SOC) pointing oppositely in *K/K'* valleys (inset to Fig. 3b). Hence, a Zeeman field – normally the universal pair breaking mechanism in other superconductors – strongly protects the Cooper pair against the orthogonal in-plane external magnetic field. Compared with other quasi-2D Ising superconductors such as ion-gated multilayer MoS$_2$[23,24] where the electronic wave function is confined in the uppermost layer and Ising pairing is protected by SOC ~ 6 meV, a much heavier transition metal in WS$_2$ monolayer provides a five times larger SOC[32] (~30 meV) suggesting even higher in-plane upper critical field $B_{c2}$ due to stronger Ising protection.

Guided by the established phase diagram shown in Fig. 2a, we induced superconductivity in another sample B (inset of Fig. 3a, optimal $T_c$ = 3.15 K using 50% of $R_N$ criteria) and measured $B_{c2}$ of three representative states at the left side (upper panel, Fig. 3a), peak (middle panel), and right side (bottom panel) of the dome, respectively. For all measurements, the 50% of $R_N$ standard was used for determining $B_{c2}$. Near the left QCP, where relatively weak electric fields cause a minimum Rashba effect[23], the strongest Ising protection was measured (Fig. 3a, upper panel): $T_c$ = 1.54 K only shifted by $\Delta T_c$ = 0.08 K for an in-plane field as high as $B$ = 35 T. Following mean-field theory[23] (also Methods), we can estimate the contribution of Zeeman type SOC and Rashba effect to the enhancement of $B_{c2}$ (relative to the Pauli limit $B_p$ = 1.86 $T_c$,); the comparison to the experimental data (solid red points) is shown in Fig. 3b (red line) Considering Rashba effect created by the strong field of ionic gating in the order of 0.7 meV, which is similar to that found near the left QCP of MoS$_2$[23], the obtained Zeeman type SOC equals to 30 meV, consistent with theoretical calculation. Neglecting Rashba contribution, our data sets the lower bound of 19.5 meV for Zeeman SOC. Because of the very large $B_{c2}$, for the states at the peak (Fig. 3a, middle panel) and right side (Fig. 3a, bottom panel) of the dome, the change of $T_c$ at maximum $B = 12$ T is below the measurement accuracy for extracting quantitative values of the SOC.

In Fig. 3b, we compared the $B_{c2}/B_p$ of other superconductors where the very large $B_{c2}$ violating significantly the Pauli limit was also observed. For the gated monolayer WS$_2$, the $B_{c2}/B_p$ is at least as large as in the UCoGe[33] and the sub-monolayer Pb film[34] protected by triplet pairing and Rashba effect, respectively. It is much larger than other recently discovered Ising-protected superconductors such as multilayer MoS$_2$[23] and monolayer NbSe$_2$[20]. It is noteworthy that the Zeeman SOC in the NbSe$_2$ (~70 meV, valence band) is even larger than that in both MoS$_2$ (~ 6 meV) and WS$_2$ (~30 meV), whereas the effect of protection in the monolayer NbSe$_2$ and the bulk (LaSe)$_{1.14}$(NbSe$_2$)[35] (approaching decoupled monolayers at low temperature) merely approaches $B_{c2}$ of the gated MoS$_2$. A similar mismatch also appears in the TaS$_2$(Py)$_{0.5}$ where pairing is also in valence band[36]. This discrepancy could be influenced by a competing charge density wave (CDW) phase and the contributions from a spin degenerate $\Gamma$ point[37], which might effectively weaken the strong Ising pairing formed at the $K/K'$ valleys of NbSe$_2$ and TaS$_2$.

The highly unexpected feature is the reversible re-entrance into a strongly insulating state at high $V_{eff}$. The signature of this fully insulating state has been observed in many systems manifesting as mobility peaks[2,38–40]. Especially, in ion-gated devices, universal conductivity peak under strong gating has been observed in a narrow band (rubrene[38] and ReS$_2$[41]) and wide band materials (silicon inversion layer[39] and multilayer WSe$_2$[40]). This material-independent characteristic may rule out many exotic mechanisms such as the opening of Mott[42], CDW[43], Kondo[44] gaps or strong correlation[45,46] that might cause insulating behaviors as well. To address this universal insulator, we focus on the common fact that all the above examples involve ionic gating.

A direct comparison in Arrhenius plots (Fig. 4a) yields similar behavior at high temperature for the insulator (left to the dome) and the most insulating states of re-entrant insulator (right to the dome). Without knowledge of underlying transport mechanisms, we extract characteristic energy scale using activation model. As shown in Fig. 4b, although the two energy scales are similar in magnitude, the signs of their dependencies on the $V_{eff}$ are opposite. The temperature dependence of $n_{Hall}$ also confirms insulating behavior at large $V_{eff}$, manifesting as freeze-out of carriers for a large range of gate voltages including the whole superconducting range ($V_{eff}$ from 0.8 to 1.6 kV, Fig. 4c). In contrast, a nearly temperature independent $n_{Hall}$ is observed only in the qM region near the left QCP at low $V_{eff}$.

To account for the anomalous re-entrant insulator, we propose a scenario of a $V_{eff}$ dependent band variation as sketched in Fig. 4d. In our truly 2D system, weak out-of-plane screening[2] exposes induced carriers directly to the potential of ions lying on the surface. The localization effect depends on the distance $l$ between induced carriers and charge centers of the cations (Fig. 4d). At low $V_{eff} \ll$ 1.1 kV, ions accumulated by weak electric field create a uniform average potential, while at large $V_{eff} \gg$ 1.1 kV, the discreteness of ions at a reduced $l$ due to the strong electric field can no longer be averaged out by characteristically weak screening in 2D. Hence, the increased randomness enlarges the band tail (Fig. 4e) where more carriers localize, reducing the number of free carriers available for band transport. In the high gating limit, every induced carrier is localized/bounded on site by the potential of an adjacent ion, *i.e.*, forming electrons-cation pair mimicking hydrogen impurity model (Fig. 4e, right panel). Strongly localized electrons in the re-entrance regime would form impurity band reducing Fermi level with the increase of gate voltage. When the temperature is sufficiently low, instead of thermal activation to the conduction band, hopping between localization centers is more plausible to dominate the transport.

With this physical picture in mind, we can understand the shift in maxima between $n_{Hall}$ (10K) and $T_c$ (Fig. 2) in the phase diagram. Starting from $V_{eff} \sim$ 1130 V where $T_c$ peaks, $n_{Hall}$ decreases from 160 to 10 K with increasing $\Delta n_{Hall}$, due to the increasing localization at higher $V_{eff}$. The localization of induced carriers forms neutralized Coulomb traps (Fig. 4d, dashed line) as short-range scattering centers causing a suppression of $T_c$ at a relatively high carrier density and a decrease in mobility (Fig.

2 lower panel). As a result, the interplay between carrier and disorder shapes the superconducting dome versus $V_{eff}$.

Considering that electrostatic gating by polarising dielectrics (polarised dipole) and ionic media (cation–anion pair) are equivalent, the competing quantum phases induced in the present study represents the high field limit in respect of superconducting domes reported previously[7,27,47–49] (Fig. S5), where either a very strong field effect or an isolated monolayer crystal is missing. In this truly 2D system, localized states can now be easily formed because of the slow decay of ion potential as $\sim 1/r^3$ at a distance $r$ from the ion indicating that any disorder in the potential landscape has a long-range effect[2]. In contrast, in quasi-2D cases, gate induced carriers always extend to a finite thickness; strong gating populates multiple sub-bands causing crossover to 3D, which enhances screening thus reduces carrier localization. In this sense, the present mechanism of quantum phases evolution provides a clear understanding of the power law/logarithmic correction in the normal state of ion-gated rubrene[38] and silicon[39] as the precursor of re-entrance and universally observed superconducting dome in gated $KTaO_3$[47], multilayer $MoS_2$[27], $ZrNCl$[48] and $TiSe_2$[49], as well as $LaAlO_3/SrTiO_3$[7,31] interface, where accessing the right QCP and the insulating state subsequent to the superconducting dome are prohibited by the enhanced screening in these quasi-2D systems.

**Figures and captions**

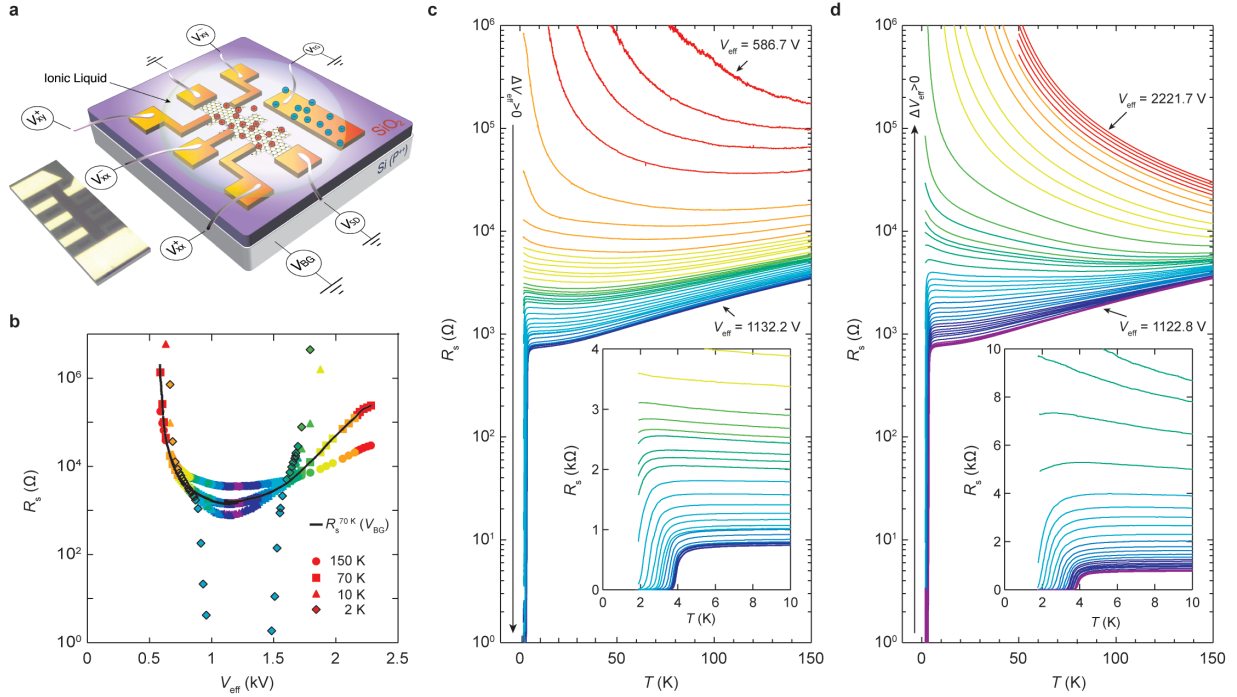

**Figure 1. Electrical transport of ion-gated monolayer WS$_2$.** **a**, Schematic of measurement set-up with both ion liquid $V_{TG}$ and solid back gates $V_{BG}$. Inset: Optical image of sample etched into standard Hall bar geometry. **b**, Transfer characteristics measured by scanning $V_{BG}$ at 70 K with various $V_{TG}$ were concatenated, as indicated by a black line. The origin of effective gate voltage $V_{eff}$ was extrapolated using gate dependence of Hall carrier density (Methods, Fig. S2). Square resistance $R_s$ at typical temperatures (150 K: circle, 70 K: square, 10 K: triangle, 2 K: diamond) are shown for many different $V_{eff}$ to reveal the evolution from the insulator, superconductor to re-entrant insulator transitions. Each color represents a specific $V_{TG}$. **c–d**, Temperature dependent $R_s$ are plotted for regimes before (**c**) and after (**d**) the peak of the superconducting dome, where each curve corresponds to one $V_{eff}$ in **b**. Insets: details around the superconducting-insulating transition in linear scale.

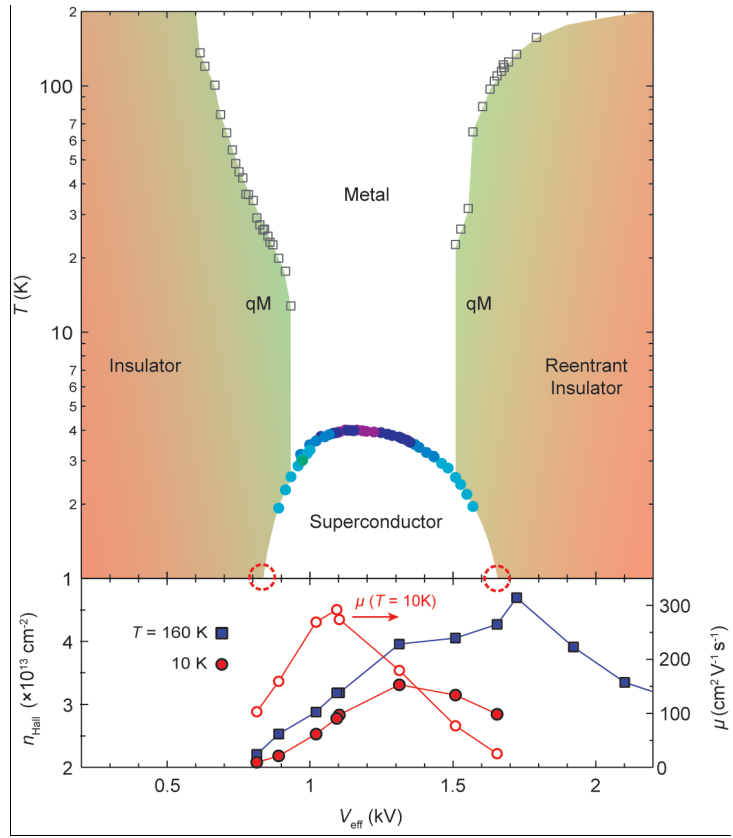

**Figure 2. Phase diagram of monolayer WS$_2$ and critical scaling of quantum phases. Upper panel**, Superconducting critical temperatures $T_C$ are plotted to the left axis (solid circle) as a function of the effective back gate. Quasi-metal (qM) regime is bounded by metal-insulator crossover temperature (empty square). The initiation and suppression of the superconducting dome are indicated by dashed circles. **Lower panel**, Hall carrier density measured at 160 and 10 K are plotted to the left axis; Hall mobility at 10 K to the right axis.

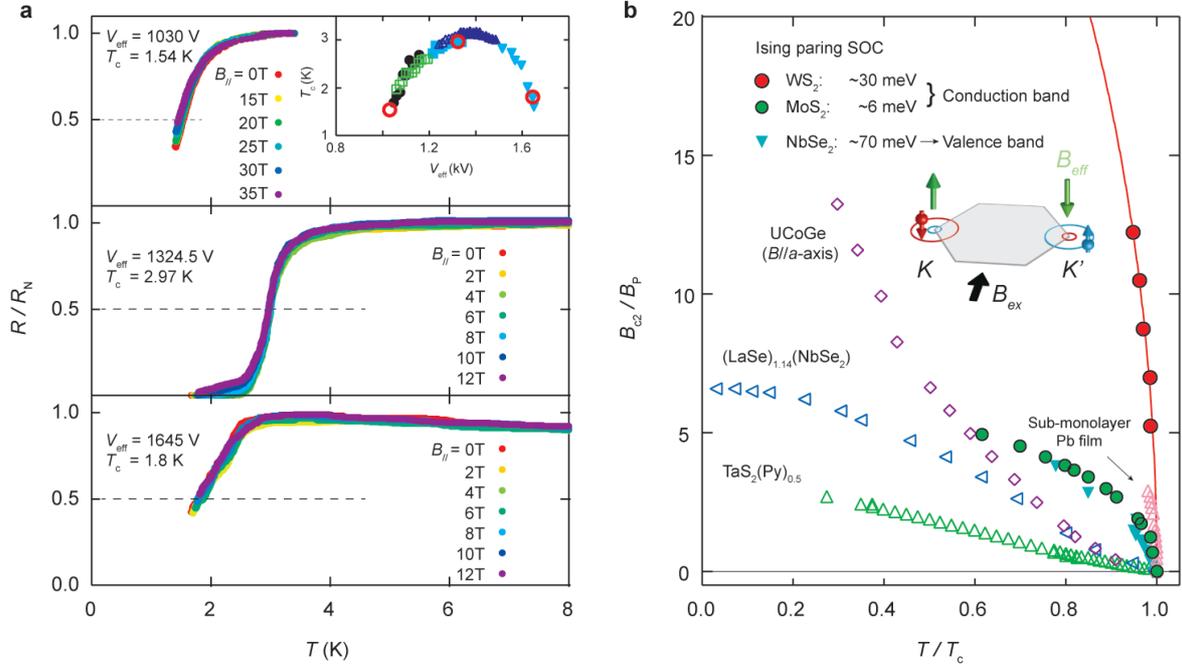

**Figure 3. Ising paring over the entire superconducting dome. a**, Upper critical field $B_{c2}$ was measured on the left side (upper panel), the peak (middle panel) and right side (bottom panel) of the superconducting dome (inset to **a**). Superconducting critical temperature $T_c$ is defined by 50% of normal state resistance denoted by dashed line. Each state is highlighted by an empty circle in the dome. **b**, Normalized $B_{c2}$ with respect to Pauli limit in $WS_2$ is denoted by the solid red circle, which exceeds that of many well-known superconductors with high $B_{c2}$ including TMDs, triplet pairing, and monolayer Pb film. Inset: schematic of Zeeman-type effective magnetic fields (green arrows) with alternating directions in K/K' valleys in a hexagonal Brillion zone, which stabilize electron spins (red/blue denotes spin up/down) in a Cooper pair against external in-plane magnetic field $B_{ex}$.

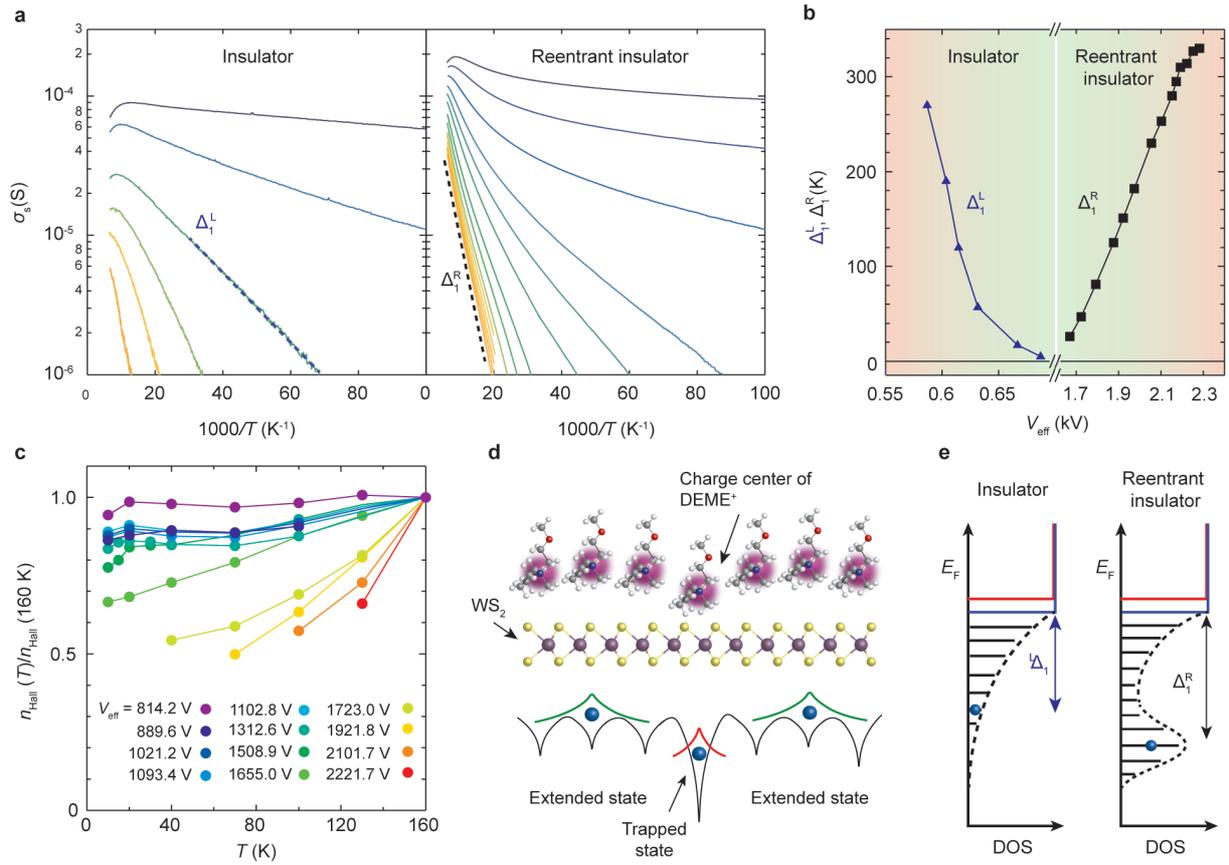

**Figure 4. Re-entrant insulator induced in monolayer WS$_2$ by ionic gating. a**, Arrhenius plot of conductance defined as $1/R_s$ in insulating (left panel) and re-entrant insulating (right panel) regimes. The characteristic energy scales are extracted in terms of thermal activation transport (dashed lines). In the right panel, the long tail at low temperature may suggest complicated hopping mechanisms (Fig. S9) along with the increasing $V_{eff}$. **b**, Extracted characteristic energy are plotted as a function of effective gate voltages. Black squares and blue triangles correspond to re-entrant insulator and band insulator. **c**, Normalized Hall carrier density at various gate voltages as a function of temperature. Free carriers freeze out during cooling down in the re-entrant insulating regime (red) while the carrier concentration almost remains constant in the metallic regime (blue and purple). **d**, Schematics of electron (blue sphere) localization in the Coulomb traps (black curve) due to the poorly screened cations (organic molecular DEME$^+$, the positive charge center is highlighted by a solid ball in purple) in proximity to monolayer WS$_2$ film. **e**, Representation of the density of states (DOS) as a function of energy (E) in the insulating phase (left of the dome) and the re-entrant insulator (right of the dome), in both of which a disorder potential results in a localised band tail below the spin-split conduction band (red and blue denote spin up and down, respectively). The insulating side has a low density of localization centers (left panel). Whereas, overlapping of high-density localized states on re-entrant insulating side plausibly forms an impurity band.